\documentclass{icrc2009}

\usepackage{graphicx}   
\usepackage[caption=false]{caption}    
\usepackage[font=footnotesize]{subfig} 
\usepackage{fixltx2e}
\usepackage{url}

\newcommand{\shorttitle}[1]%
{\markboth{Proceedings of the 31\MakeLowercase{$^{st}$} ICRC, {\L}\'{o}d\'{z}
2009}{#1} }
\newcommand{\etal}{\MakeLowercase{\textit{et al. }}} 

\begin{document}
\title{Cosmic-ray electrons, synchrotron and magnetic fields in the Galaxy}

\author{\IEEEauthorblockN{E. Orlando\IEEEauthorrefmark{1},  A.W. Strong\IEEEauthorrefmark{1},  I.V. Moskalenko\IEEEauthorrefmark{2}, T.A. Porter\IEEEauthorrefmark{3}, G. Johannesson\IEEEauthorrefmark{2}, S. W. Digel\IEEEauthorrefmark{4}
                          }
                            \\

\IEEEauthorblockA{\IEEEauthorrefmark{1}Max-Planck-Institut f\"ur extraterrestrische Physik, Postfach 1312, D-85741 Garching, Germany}
\IEEEauthorblockA{\IEEEauthorrefmark{2}W. W. Hansen Experimental Physics Laboratory,\\
Kavli Institute for Particle Astrophysics and Cosmology,\\
Stanford University, Stanford, CA 94305 
}
\IEEEauthorblockA{\IEEEauthorrefmark{3}Santa Cruz Institute for Particle
Physics, University of California, 1156 High Street, Santa Cruz, CA
95064}
\IEEEauthorblockA{\IEEEauthorrefmark{4}Kavli Institute for Particle Astrophysics and Cosmology,\\
SLAC National Accelerator Laboratory, Stanford, CA 94025}

}

\shorttitle{E. Orlando \etal Cosmic-ray electrons, synchrotron and magnetic fields in the Galaxy}
\maketitle

\begin{abstract}
The cosmic-ray propagation code GALPROP has recently been enhanced in the areas of Galactic magnetic field and synchrotron radiation. A full 3D treatment is now implemented. We describe these enhancements and show applications to the interpretation of non-thermal radio surveys and gamma rays, to probe the Galactic cosmic-ray electron spectrum and magnetic fields.
\end{abstract}

 
\section{Introduction}
Cosmic-ray electrons and positrons produce synchrotron radiation by gyrating in the magnetic field of our Galaxy. The same electrons are also partly responsible for the Galactic diffuse gamma-ray emission, in the hard X-ray to gamma-ray energy range \cite{Porter2008}. Since cosmic rays, gamma rays, synchrotron emission and Galactic magnetic field are interrelated, combining such observations constrains cosmic-ray electrons and components of the interstellar medium (ISM) responsible for
the diffuse emission over this broad energy range.

At the conference we will present a study of the synchrotron emission in order to constrain magnetic field models of the Galaxy, using the knowledge from the gamma-ray and electron data of the Fermi telescope.

We use the GALPROP code for the calculation of Galactic cosmic-ray propagation (\cite{strong1998}, \cite{moska1998}, \cite{SMR00},  \cite{moska2000}, \cite{strong2004} and \cite{Strong2007}). 
The code is developed to reproduce direct cosmic-ray measurements, gamma rays and
now synchrotron radiation data. More details of the GALPROP code and its developments are given in \cite{Strong2009} and in the next paragraph.

\section{GALPROP developments}
Synchrotron emission depends on the model of the magnetic field. Hence, an implementation of a proper 3D model of the Galactic magnetic field is fundamental to have a good description of the diffuse synchrotron radiation and its latitude and longitude distribution.
Until now, a simplified magnetic field model in 2D was implemented in
GALPROP, using only
a random component with an exponential decrease
in Galactocentric radius and height above the plane characterised by a scale
length in each dimension.
With the availability of excellent radio continuum surveys from tens of MHz to tens
of GHz, including the WMAP satellite data, more sophistication is desirable.
Hence GALPROP has been extended to include a general model of the B-field and the
associated synchrotron emission. We have introduced full 3D models for both
regular and random magnetic fields. The routines have been adapted to compute the emission in 3D; with the previous 2D formalism the asymmetries in the synchrotron emission were not described.

At present the Galactic magnetic field is not well known. Hence many models with different configurations can be found in the literature. For our study we chose some of them as a starting point and made our own choice of the parameters.
The formulation of the regular component of the magnetic field introduced in GALPROP follows a spiral form. 
The regular component is modelled as a vector field $\bf{B}_{reg}$, the random field
as a scalar field  $B_{ran}(x,y,z)$, with the option
also to represent it  as a vector in case explicit modelling of random orientations
is required.

Skymaps, longitude and latitude profiles at a variety of energies/frequencies, as well as spectral index plots are produced as outputs so that the diffuse emission for different models of the magnetic field 
can be compared with data.

With the present version of GALPROP only the total synchrotron intensity can be compared with data; polarized emission will be implemented in future work.

Electrons and positrons lose energy by inverse Compton and synchrotron radiation, and the latter is included
in GALPROP self-consistently using the total field of the adopted model.
For the spectrum of particles (here electrons and positrons) computed by GALPROP at
all points on the 3D grid, 
 we integrate over particle energy and line-of-sight to get the synchrotron
intensity for the regular and random fields.
The resulting synchrotron skymaps for a user-defined grid of frequencies are output
by GALPROP either in Cartesian (l,b) or in  HEALPix isopixelisation format\footnote{http://healpix.jpl.nasa.gov}.
The emissivity as seen by an observer at the solar position is also output as a
function of $(x,y,z,\nu)$.

\section{Radio surveys}
The aim is to describe both spectral and spatial properties of the
synchrotron sky, and not the fine angular details.
For the spatial distribution the most useful data are the Haslam 408 MHz survey (\cite{haslam}),
since it has full sky coverage, a well-established calibration and zero
level, and is dominated by synchrotron emission.
The other data used in this paper are from
22 MHz: DRAO survey (\cite{roger});
150 MHz: \cite{landecker}; 1420 MHz: \cite{reich86} and \cite{reich01}. 
For WMAP frequencies we used the spectral-index maps generated from WMAP polarized data by \cite{Miville}, scaling from
408 MHz to give total synchrotron intensity.

\section{Procedure}
We have implemented many 3D models of the Galactic magnetic field in GALPROP in order to calculate the synchrotron emission from the Galaxy. 
Our approach is to use the most up to date models of the magnetic field from the literature (e.g. \cite{Miville}, \cite{han2006}, \cite{prouza2003}, \cite{sun}, \cite{TT}) and adjust the parameters in order to reproduce at the same time gamma-ray and radio observations.

In future we will use a distribution of cosmic-ray sources and an electron spectrum that fits the
Fermi gamma-ray spectrum and electron measurements. Then the value of total magnetic field will be adjusted to fit the synchrotron 408 MHz map. We will compare synchrotron latitude and longitude profiles with the available radio surveys.
The best Galactic magnetic field model that fits simultaneously cosmic-ray, gamma and radio data will be presented at the conference. Results will be given using the cosmic-ray electron spectrum measured by Fermi \cite{fermi_el}. 

\section{Results}

Here we show an example of the outputs that can be obtained with GALPROP for a given model of the magnetic field. In this case the electron spectrum used was the so called conventional model defined in  \cite{strong2004}, based on pre-Fermi electron measurements.  

The regular Galactic magnetic field used in this example is an axisymmetric logarithmic spiral model with a local intensity of 6 $\mu$G and a pitch angle of 8.5$^\circ$. The halo magnetic field is given by the \emph{z}-dependence taken from \cite{Miville}, but the parameters are adjusted to fit the latitude profiles of the synchrotron emission. The disk random field was taken as 3 $\mu$G, constant in Galactocentric distance, with a halo component decreasing in \emph{z}
with the same law as the regular one. Toroidal and dipolar magnetic field components such as in \cite{sun} are also added. 

\begin{figure}[h]
\centering
\includegraphics[width=0.5\textwidth, angle=0] {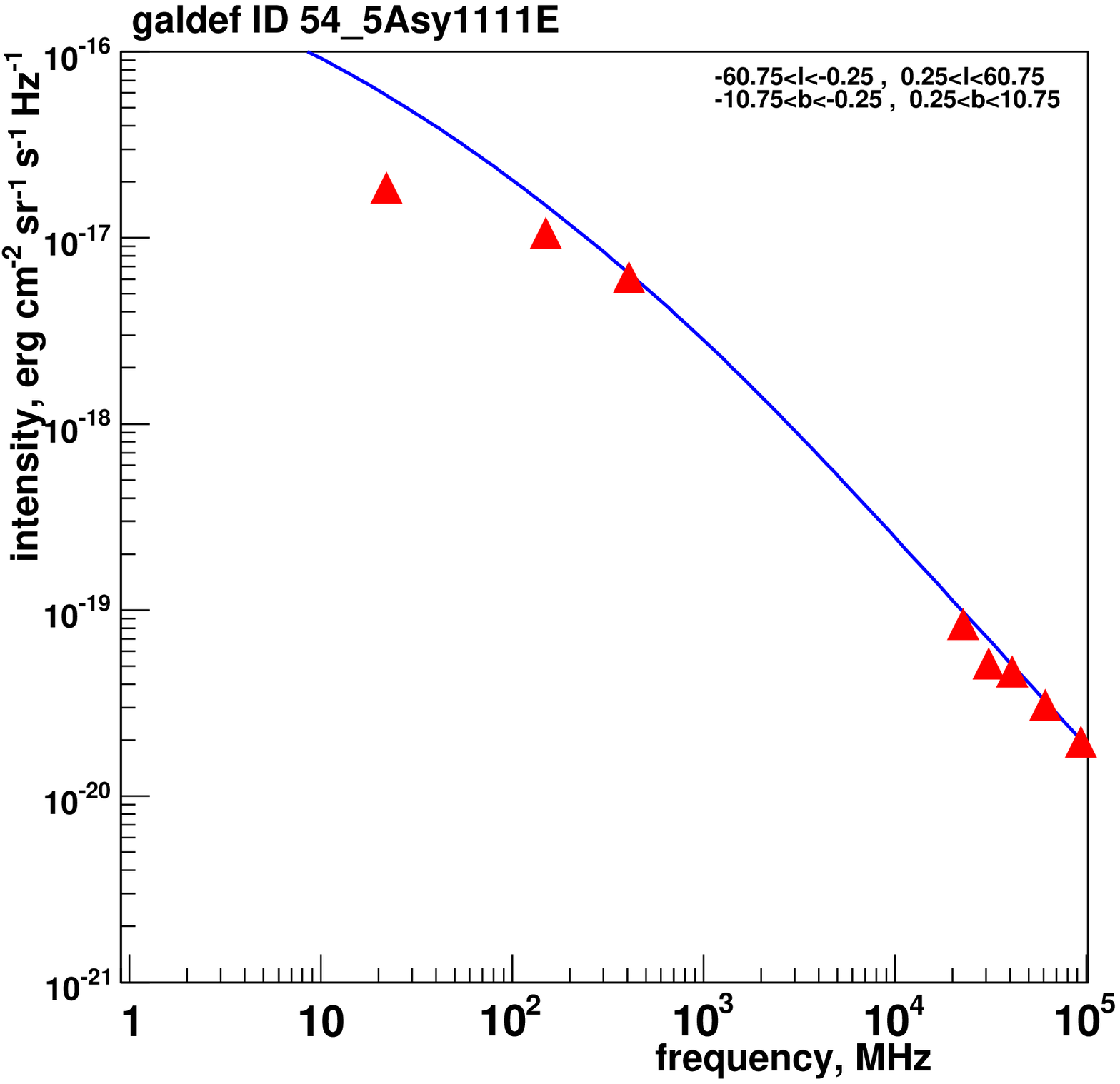}\\
\includegraphics[width=0.5\textwidth, angle=0] {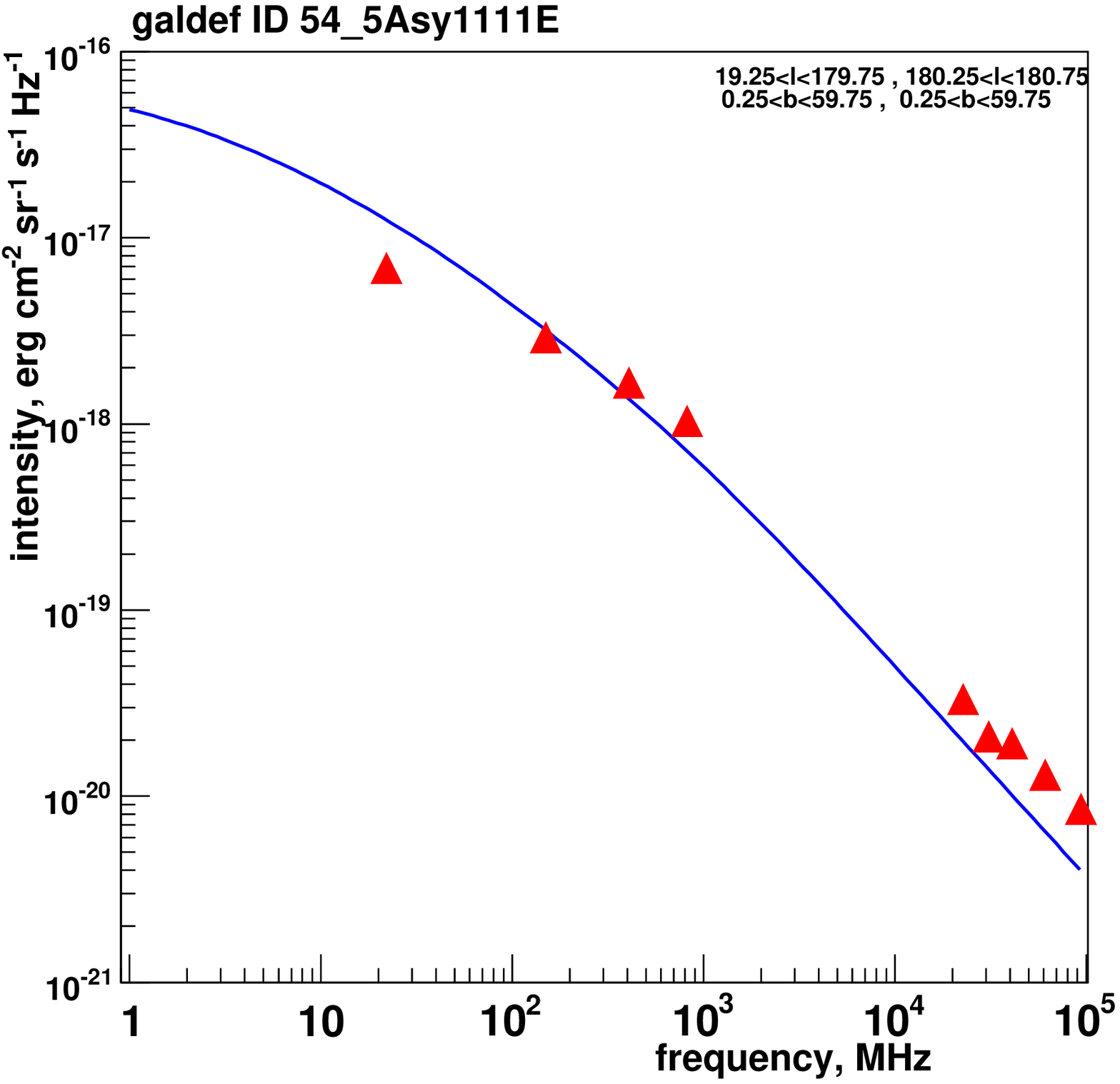}
 \caption{Synchrotron spectra for the inner Galaxy (left)  (300$^\circ$$<$l$<$60$^\circ$,  $\vert$b$\vert$$<$10$^\circ$) and for the region 20$^\circ$$<$l$<$180$^\circ$,  $\vert$b$\vert$$<$60$^\circ$ (right). Blue lines are obtained with the GALPROP code, while red points are radio surveys. The data are from
22 MHz: DRAO survey (\cite{roger});
150 MHz: \cite{landecker}; 408 MHz: \cite{haslam}; 1420 MHz: \cite{reich86} and \cite{reich01}; 
WMAP: \cite{Miville}.
}
\label{fig1}
 
\end{figure}

\begin{figure*}[h]
\centering
\includegraphics[width=0.5\textwidth, angle=0] {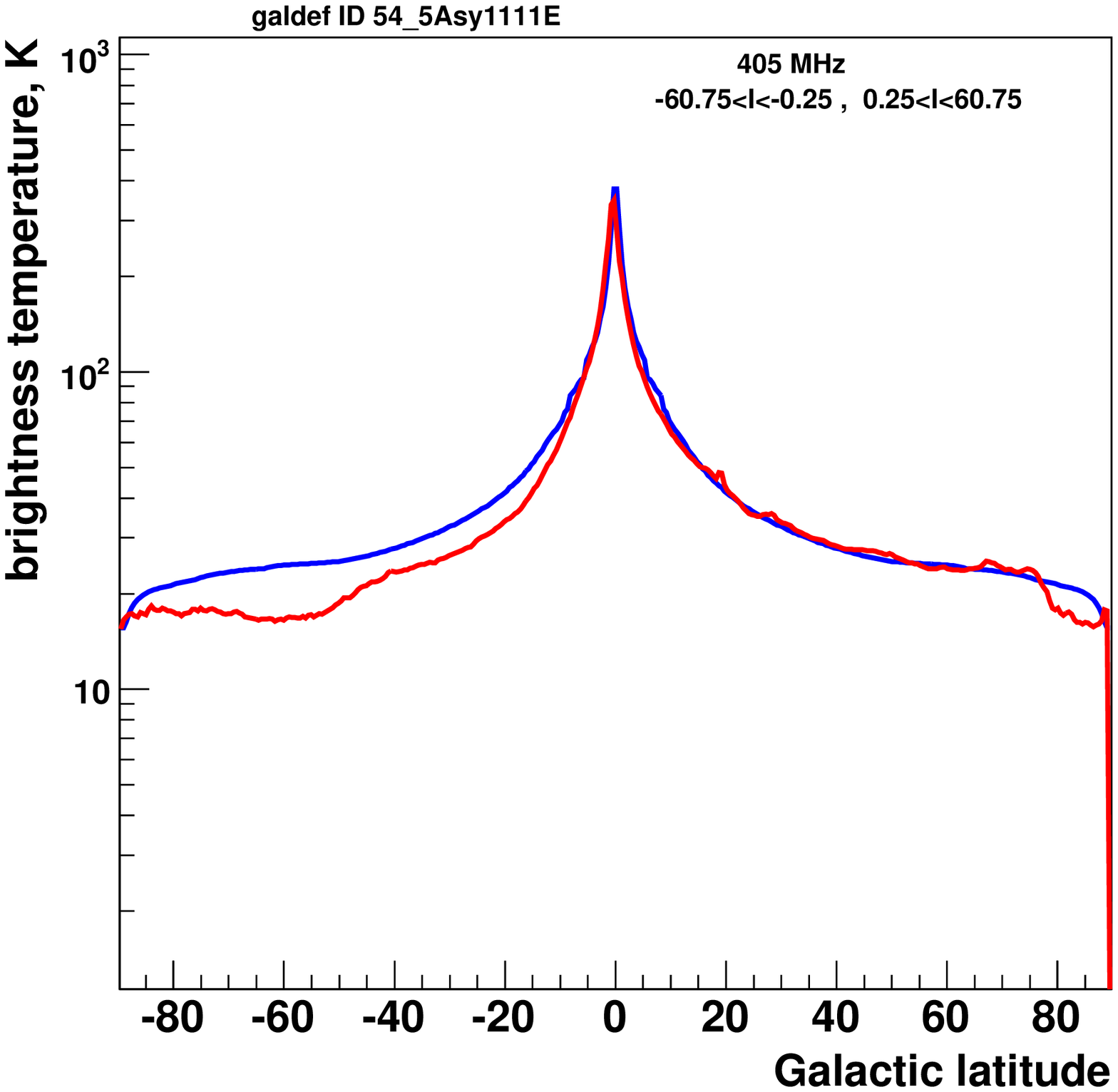}\\
\includegraphics[width=0.5\textwidth, angle=0] {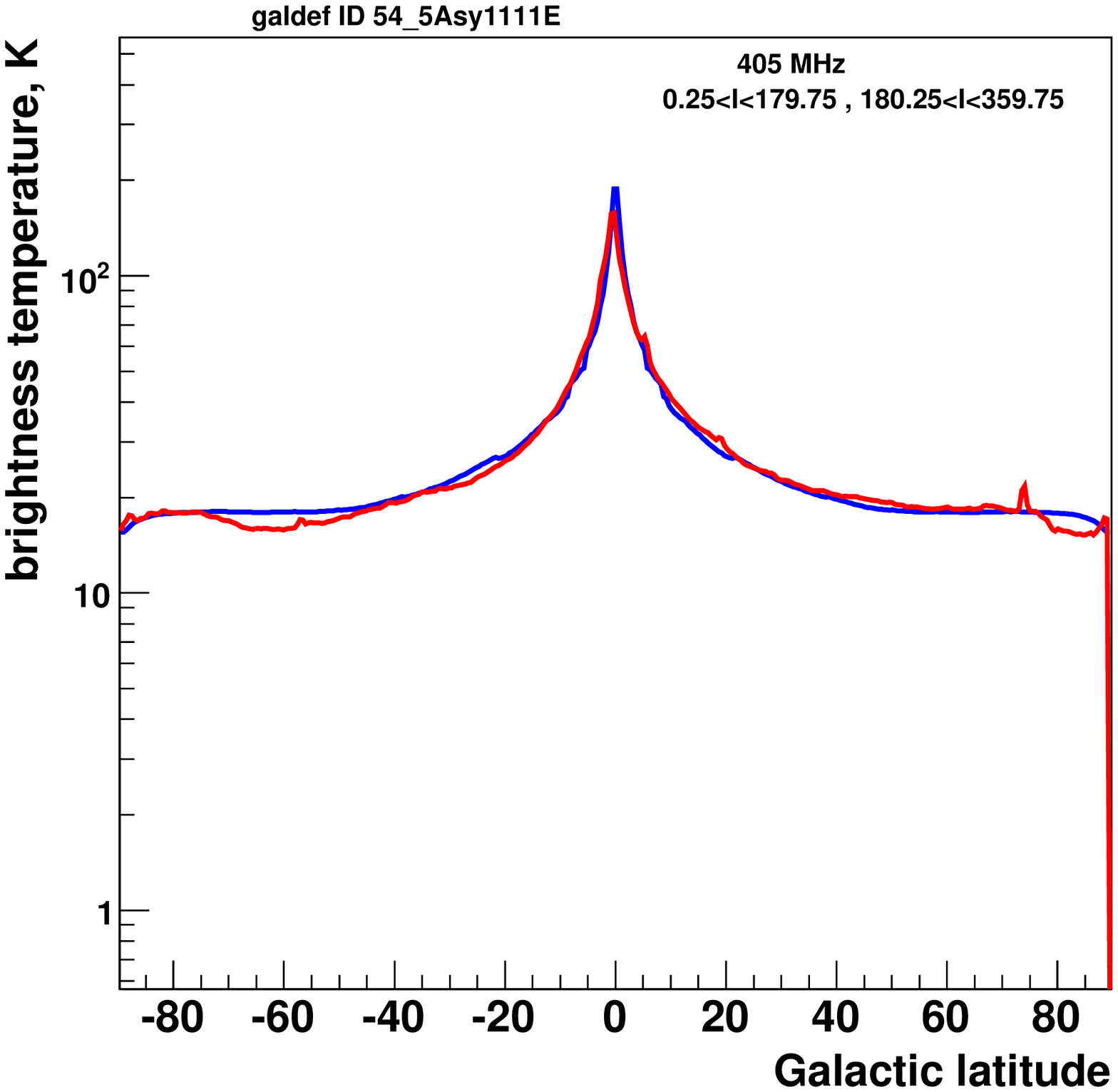}\\
\includegraphics[width=0.5\textwidth, angle=0] {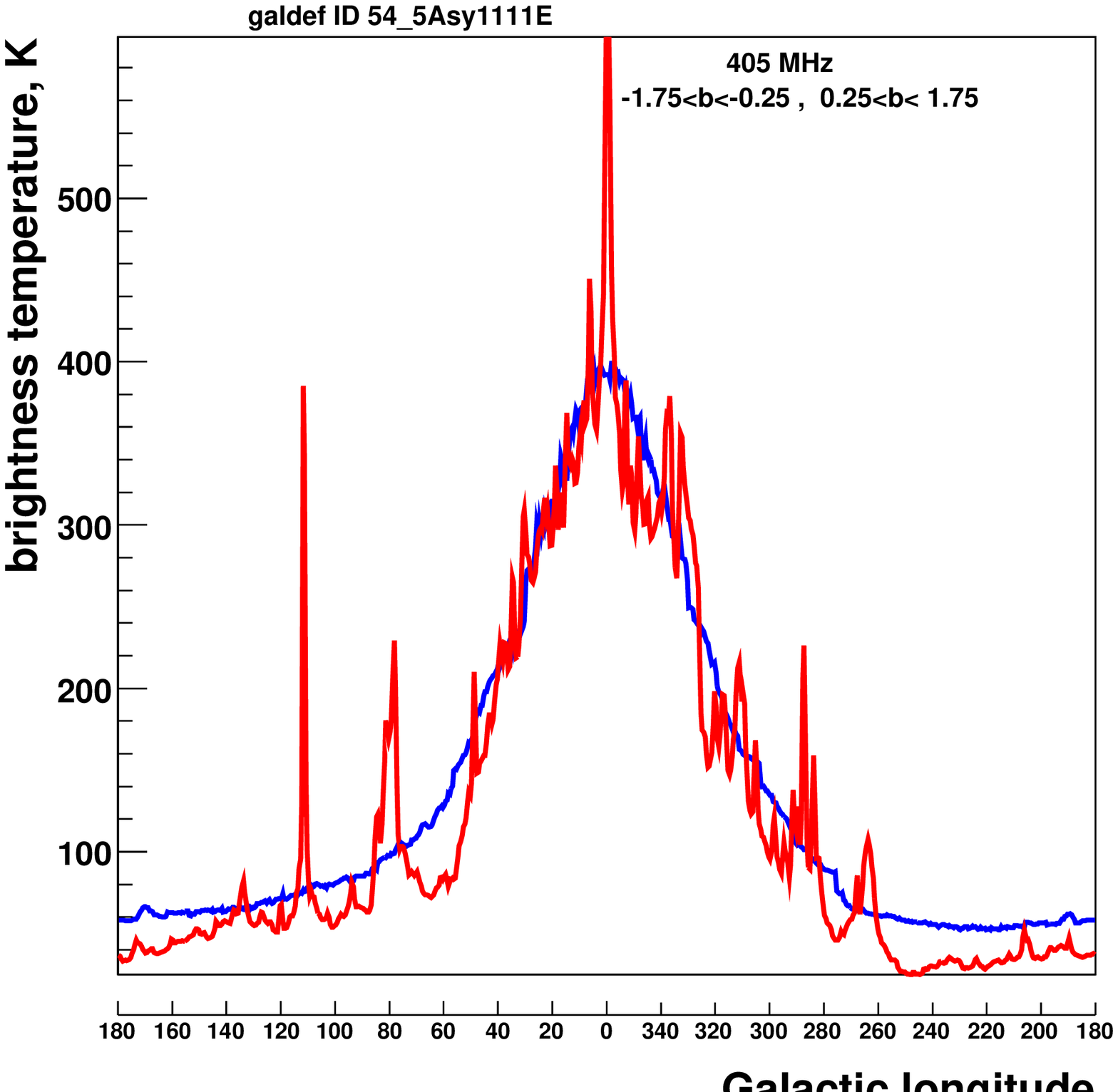}\\
  \caption{Intensity profiles of synchrotron emission at 408 MHz in latitude for the inner Galaxy, $\vert$l$\vert$ $<$60$^\circ$, and the whole Galaxy, and in longitude for the Galactic plane region. Blue lines are obtained with the GALPROP code, while red lines are data from \cite{haslam}. The zero level is corrected as described in \cite{reich88} including subtraction of the CMB and extragalactic background.}
  \label{fig2}
 
\end{figure*}

Synchrotron spectra for the inner and the northern Galaxy obtained with GALPROP are shown in Fig.\ref{fig1} compared with radio surveys. The spectrum is reproduced except at the lowest frequencies where
absorption is evident, which is not yet included in the model. Resulting latitude and longitude profiles of the synchrotron emission at 408 MHz are plotted in Fig. \ref{fig2}, compared with data from \cite{haslam}. The general form of the profiles are already well fitted with this preliminary model.

\begin{thebibliography}{99}
 \bibitem{Porter2008} Porter, T.~A., 
Moskalenko, I.~V., Strong, A.~W., Orlando, E., 
\& Bouchet, L.\ 2008, ApJ, 682, 400 

 \bibitem{strong1998} Strong, A.~W., \& Moskalenko, I.~V.\ 1998, ApJ, 509, 212 
\bibitem{moska1998} Moskalenko, I.~V., \& Strong, A.~W.\ 1998, ApJ, 493, 694 
\bibitem{SMR00}
  Strong, A. W., Moskalenko, I. V., \& Reimer, O.\ 2000, ApJ, 537, 763
\bibitem{moska2000} Moskalenko, I.~V., \& Strong, A.~W.\ 2000, ApJ, 528, 357 
\bibitem{strong2004} Strong, A.~W., 
Moskalenko, I.~V., \& Reimer, O.\ 2004, ApJ, 613, 962 
\bibitem{Strong2007}
  Strong, A. W., Moskalenko, I. V., \& Ptuskin, V. S.\ 2007, Ann. Rev. Nuc. Part. Sci., 57, 285
  
\bibitem{Strong2009}  A. W. Strong, I. V. Moskalenko, T. A. Porter, G. Johannesson, E. Orlando, S. W. Digel, these proceedings

 

\bibitem{haslam} Haslam, C.~G.~T., Salter, C.~J., Stoffel, H., \& Wilson, W.~E.\ 1982, A\&AS, 47, 1 

\bibitem{roger} Roger, R.~S., Costain, C.~H., Landecker, T.~L., \& Swerdlyk, C.~M.\ 1999, A\&AS, 137, 7 
\bibitem{landecker} Landecker, T.~L., \& Wielebinski, R.\ 1970, Australian Journal of Physics Astrophysical Supplement, 16, 1 

\bibitem{reich86} Reich, P., \& Reich, W.\ 1986, A\&A, 63, 205 


\bibitem{reich01} Reich, P., Testori, J.~C., \& Reich, W.\ 2001, A\&A, 376, 861 




\bibitem{Miville} Miville-Desch{\^e}nes, M.-A., Ysard, N., Lavabre, A., Ponthieu, N., Mac{\'{\i}}as-P{\'e}rez, J.~F., Aumont, J., \& Bernard, J.~P.\ 2008, A\&A, 490, 1093 
\bibitem{han2006} Han, J.~L., Manchester, 
R.~N., Lyne, A.~G., Qiao, G.~J., \& van Straten, W.\ 2006, ApJ, 642, 868 
\bibitem{prouza2003} Prouza, M., \& {\v S}m{\'{\i}}da, R.\ 2003, A\&A, 410, 1 

\bibitem{sun} Sun, X.~H., Reich, W., Waelkens, A., \& Ensslin, T.~A.\ 2008, A\&A, 477, 573 
\bibitem{TT} Tinyakov, P.~G., \& Tkachev, I.~I.\ 2002, Astroparticle Physics, 18, 165 

\bibitem{fermi_el} Abdo, A. A., et al. \ 2009, Phys. Rev. Lett. 102, 181101
\bibitem{reich88} Reich, P., \& Reich, W.\ 1988, A\&AS, 74, 7 





 
\end{thebibliography}
\end{document}